\documentclass[conference]{IEEEtran}
\IEEEoverridecommandlockouts
\usepackage{eurosym}
\usepackage{amssymb}
\usepackage{amsmath}
\usepackage{amsfonts}
\usepackage{float}
\usepackage{graphics}
\usepackage[tight,footnotesize]{subfigure}
\usepackage{rotating}
\usepackage{algorithm}
\usepackage{enumerate}
\usepackage{cite}
\usepackage{xcolor}
\usepackage{color,soul}
\usepackage[colorinlistoftodos]{todonotes}
\usepackage{lipsum}
\usepackage{cases}
\usepackage{comment}
\usepackage{algpseudocode}
\usepackage{resizegather}

\usepackage{multirow}
\usepackage{threeparttable}

\usepackage{mathtools}
\usepackage{makecell}

\DeclareMathOperator{\var}{Var}

\DeclareMathSizes{10}{9}{7}{5}
\newtheorem{theo}{Theorem}
\newtheorem{prop}{Proposition}
\newtheorem{theorem}{Theorem}

\newtheorem{corollary}{Corollary}

\newtheorem{definition}[theorem]{Definition}

\newtheorem{lemma}{Lemma}

\newtheorem{example}{Example}

\newcommand{\bs}{\boldsymbol{\mathrm{s}}}
\newcommand{\br}{\boldsymbol{\mathrm{r}}}
\newcommand{\be}{\boldsymbol{\mathrm{e}}}

\newcommand{\bu}{\boldsymbol{\mathrm{u}}}
\newcommand{\bt}{\boldsymbol{\mathrm{t}}}
\newcommand{\bA}{\boldsymbol{\mathrm{A}}}
\newcommand{\buone}{\bA^{T}\br+\be_{1}}

\newcommand{\coloneqq}{:=}

\def\BibTeX{{\rm B\kern-.05em{\sc i\kern-.025em b}\kern-.08em
    T\kern-.1667em\lower.7ex\hbox{E}\kern-.125emX}}

\begin{document}

\title{On the Maximum Toroidal Distance Code for Lattice-Based Public-Key Cryptography
}

\author{\IEEEauthorblockN{Shuiyin  Liu}
\IEEEauthorblockA{{Cyber Security Research and Innovation Centre} \\
{Holmes Institute}\\ {Melbourne, VIC 3000, Australia} \\
Email: SLiu@Holmes.edu.au}
\and
\IEEEauthorblockN{Amin Sakzad}
\IEEEauthorblockA{
%{Department of Software Systems $\&$ Cybersecurity} \\
{Faculty of Information Technology}\\
{Monash University}\\ {Melbourne, VIC 3800, Australia} \\
Email: Amin.Sakzad@monash.edu}
}

\maketitle

\begin{abstract}
We propose a maximum toroidal distance (MTD) code for lattice-based public-key encryption (PKE). By formulating the encryption encoding problem as the selection of $2^\ell$ points in the discrete $\ell$-dimensional torus $\mathbb{Z}_q^\ell$, the proposed construction maximizes the minimum $L_2$-norm toroidal distance to reduce the decryption failure rate (DFR) in post-quantum schemes such as the NIST ML-KEM (Crystals-Kyber). For $\ell = 2$, we show that the MTD code is essentially a variant of the Minal code recently introduced at IACR CHES 2025. For $\ell = 4$, we present a construction based on the $D_4$ lattice that achieves the largest known toroidal distance, while for $\ell = 8$, the MTD code corresponds to $2E_8$ lattice points in $\mathbb{Z}_4^8$. Numerical evaluations under the Kyber setting show that the proposed codes outperform both Minal and maximum Lee-distance ($L_1$-norm) codes in DFR for $\ell > 2$, while matching Minal code performance for $\ell = 2$.
\end{abstract}


\section{Introduction}

Lattice-based cryptography has gained significant attention due to the threat posed by quantum computers to classical public-key schemes such as RSA and ECC. Among the lattice-based key encapsulation mechanisms (KEM), Kyber was selected as a NIST standard in 2024 \cite{NISTpqcfinal2024}, with other candidates including FrodoKEM \cite{FrodoKEM2021} and NTRU \cite{NTRU1998}. Despite strong security, a major challenge is their large ciphertexts: FrodoKEM-1344 encrypting a 32-byte secret produces $21{,}696$ bytes, and Kyber-1024 produces $1{,}568$ bytes, far exceeding classical ECC KEMs ($32$–$64$ bytes). This motivates coding and compression techniques to reduce the ciphertext expansion rate (CER) while preserving security and reliability.

To reduce the ciphertext size of lattice-based KEMs, several coding approaches have been proposed for schemes such as lattice codes for FrodoKEM \cite{FrodoCong2022}, zero divisor codes for NTRU \cite{DAWN_NTRU_2025}, and lattice codes for Kyber \cite{shuiyin2024}\cite{liu2023lattice}. These techniques reduce the decryption failure rate (DFR) by carefully encoding messages to correct decryption noise, thus allowing further compress the ciphertext. A common feature of these coding schemes is the assumption that decryption noise coefficients are independent and identically distributed (i.i.d.). The same assumption is also used in code-based KEMs such as HQC \cite{HQC2025}, which was standardized by NIST in early 2025. In practice, however, the noise coefficients are dependent, and this simplifying assumption can lead to underestimated DFR values, raising security and reliability concerns \cite{DFRAttack2019}.

To avoid the unrealistic assumption that decryption noise coefficients 
are independent, two main approaches have recently emerged for Kyber. The first approach is \emph{ciphertext packing and vertical encoding} \cite{P_kYBER2025}, where \(t\) plaintexts are packed into a single ciphertext. In this setting, the decryption decoding noise entries across the \(t\) plaintexts—also referred to as layers—are mutually independent. By encoding every \(t\) cross-layer information symbols into a single codeword, the resulting DFR analysis no longer requires independence among noise coefficients within a plaintext. The second approach is to explicitly characterize the \emph{joint probability distribution of \(\ell\) noise coefficients} arising from a polynomial product in \(R_q = \mathbb{Z}_q[x]/(x^n + 1)\) \cite{Minal_Code_Kyber2024}. By decomposing this joint distribution into a convolution of independent distributions, one can analyze the DFR of \(\ell\)-dimensional codes without invoking any independence assumptions. These approaches provide frameworks for designing coding schemes with more accurate reliability analysis in lattice-based cryptography.

Despite their advantages, these approaches have inherent limitations. 
Ciphertext packing is effective primarily when $t > 1$ independent secrets are exchanged simultaneously \cite{P_kYBER2025}. 
This, however, departs from the standard KEM setting, where only a single shared secret 
(e.g., $t=1$) is generated per encapsulation and can be only of interest in multi-recipient KEMs. 
The second approach, while theoretically sound, requires convolving $O(n)$ distinct $\ell$-dimensional joint distributions, rendering DFR evaluation and code optimization computationally prohibitive for large $\ell$. 
Consequently, existing code searches under this framework, known as \emph{Minal codes} \cite{Minal_Code_Kyber2024}, are only tractable for small dimensions, typically $\ell \leq 4$, and it remains unknown whether any codes outperform them.

In this work, we construct explicit codes for low dimensional settings ($\ell \le 8$) aimed at fast KEM implementations with tractable DFR, low CER, and low decoding complexity. The encoding task in lattice-based encryption is cast as selecting $2^\ell$ points in the discrete torus $\mathbb{Z}_q^\ell$ to maximize the minimum \emph{$L_2$-norm toroidal distance}, yielding a class of \emph{maximum toroidal distance} (MTD) codes. Unlike traditional \emph{maximum Lee distance} (MLD) codes, which rely on the $L_1$-norm, MTD codes focus on the $L_2$-norm. We target a rate of one bit per dimension, motivated by practical module- and ring-LWE KEMs with ring dimension $n = 256$, where a $256$-bit secret is exchanged. For \( \ell = 1 \), the MTD code reduces to the original Kyber encoding; for \( \ell = 2 \), it coincides with a variant of the Minal code; for \( \ell = 4 \), we present a \( D_4 \)-lattice construction achieving the largest known toroidal distance; and for \( \ell = 8 \), we obtain an optimal construction based on the \( E_8 \) lattice. Numerical evaluations under the Kyber setting show that the proposed codes outperform both Minal and MLD codes in terms of DFR for $\ell > 2$, while matching the performance of Minal codes for $\ell = 2$.

\section{System Model}
\subsection{Notation}
\emph{Rings:}  
We denote by \( R \) and \( R_q \) the rings \( \mathbb{Z}[X]/(X^n + 1) \) and \( \mathbb{Z}_q[X]/(X^n + 1) \), respectively. Throughout this work, we fix the ring dimension to \( n = 256 \). Matrices and vectors are denoted by bold upper-case and lower-case letters, respectively. The transpose of a vector \( \mathbf{v} \) or a matrix \( \mathbf{A} \) is written as \( \mathbf{v}^T \) or \( \mathbf{A}^T \). Elements of \( R \) or \( R_q \) are denoted by regular font letters, while their coefficient vectors are denoted by bold lower-case letters. All vectors are column vectors by default.

\emph{Sampling and Distributions:}  
For a set \( \mathcal{S} \), the notation \( s \leftarrow \mathcal{S} \) indicates that \( s \) is sampled uniformly at random from \( \mathcal{S} \). When \( \mathcal{S} \) denotes a probability distribution, \( s \leftarrow \mathcal{S} \) indicates sampling according to \( \mathcal{S} \). For polynomials in \( R_q \) or vectors of such polynomials, sampling is performed coefficient-wise. We write \( \var(\mathcal{S}) \) for the variance of a distribution \( \mathcal{S} \). Let \( x \) be a bit string and let \( \mathsf{Sam}(\cdot) \) denote a sampling algorithm; we write \( y \sim S \coloneqq \mathsf{Sam}(x) \) to indicate that \( y \) is sampled from distribution \( S \) using input \( x \), with output length extendable as needed. Finally, we denote by \( \beta_{\eta} = B(2\eta, 0.5) - \eta \) the centered binomial distribution over \( \mathbb{Z} \). For distributions $\mathcal{P}_X$ and $\mathcal{P}_Y$, we denote by $\mathcal{P}_X * \mathcal{P}_Y$ their convolution, where $*$ represents the convolution of distributions. For a distribution $\mathcal{P}$, we denote by $\mathcal{P}^{*t}$ the $t$-fold self-convolution of $\mathcal{P}$. We denote by $P^{\otimes t}$ the
$t$-dimensional distribution of i.i.d.\ samples drawn from $\mathcal{P}$.

\emph{Compression and Decompression (scalar quantization used in Kyber):}  
For a real number \( x \in \mathbb{R} \), the notation \( \lceil x \rfloor \) denotes rounding to the nearest integer, with ties rounded upward. The operators \( \lfloor x \rfloor \) and \( \lceil x \rceil \) denote rounding down and rounding up, respectively. Let \( x \in \mathbb{Z}_q \) and \( d \in \mathbb{Z} \) satisfy \( 2^d < q \). Following \cite{NISTpqcfinal2024}, we define the compression/decompression function as
{\setlength{\abovedisplayskip}{3pt} \setlength{\belowdisplayskip}{3pt}
\begin{align}
\mathsf{Compress}_q(x,d) &= \big\lceil (2^{d}/q)\, x \big\rfloor \bmod 2^{d}, \nonumber\\
\mathsf{Decompress}_q(x,d) &= \big\lceil (q/2^{d})\, x \big\rfloor .
\label{ComDecom}
\end{align}}

\emph{Decryption Failure Rate (DFR) and Ciphertext Expansion Rate (CER):}  
The decryption failure rate is defined as $\mathrm{DFR} =  \Pr(\hat{m} \neq m)$, where $m$ is the shared secret. A small DFR is crucial to resist decryption failure attacks \cite{DFRAttack2019}. Communication overhead is measured by the ciphertext expansion rate (CER), the ratio of ciphertext size to plaintext size.

\subsection{Kyber public key encryption (PKE)}
Let \( \mathcal{M}_{2,n} = \{0,1\}^n \) denote the message space, where each message
\( m \in \mathcal{M}_{2,n} \) is identified with a polynomial in \( R \) whose
coefficients lie in \( \{0,1\} \). We consider the public-key encryption scheme
\( \mathsf{Kyber.CPA} = (\mathsf{KeyGen}, \mathsf{Enc}, \mathsf{Dec}) \) as specified
in Algorithms~\ref{alg:kyber_keygen}--\ref{alg:kyber_dec} \cite{NISTpqcfinal2024}. The
parameters DFR, CER, and \( (q, k, \eta_1, \eta_2, d_u, d_v) \) are
summarized in Table~\ref{Kyber_Par}. The parameters \( (q, k, \eta_1, \eta_2) \)
determine the security level of Kyber, whereas \( (d_u, d_v) \) control the
ciphertext compression rate and, equivalently, the communication overhead.

\vspace{-2mm}
\begin{algorithm}[H]
\caption{$\mathsf{Kyber.CPA.KeyGen()}$: key generation}
\label{alg:kyber_keygen}
\begin{algorithmic}[1]

    \State
    $\rho,\sigma\leftarrow\left\{ 0,1\right\} ^{256}$

    \State
    $\bA\sim R_{q}^{k\times k}\coloneqq\mathsf{Sam}(\rho)$

    \State
    $(\bs,\be)\sim\beta_{\eta_1}^{k}\times\beta_{\eta_1}^{k}\coloneqq\mathsf{Sam}(\sigma)$

    \State
    $\bt\coloneqq\boldsymbol{\mathrm{As+e}}$\label{line:t}

    \State \Return $\left(pk\coloneqq(\boldsymbol{\mathrm{t}},\rho),sk\coloneqq\bs\right)$  

\end{algorithmic}
\end{algorithm}

\vspace{-5mm}

\begin{algorithm}[H]
\caption{$\mathsf{Kyber.CPA.Enc}$ $(pk=(\boldsymbol{\mathrm{t}},\rho),m\in\mathcal{M}_{2,n})$}
\label{alg:kyber_enc}
\begin{algorithmic}[1]

	\State
	$r \leftarrow \{0,1\}^{256}$
	
	%\State  $\mathrm{\boldsymbol{\mathrm{t}}}\coloneqq\mathsf{Decompress}_{q}(\boldsymbol{\mathrm{t}},d_{t})$
	
	\State
	$\boldsymbol{\mathrm{A}}\sim R_{q}^{k\times k}\coloneqq\mathsf{Sam}(\rho)$
	
	\State  $(\boldsymbol{\mathrm{r}},\boldsymbol{\mathrm{e}_{1}},e_{2})\sim\beta_{\eta_1}^{k}\times\beta_{\eta_2}^{k}\times\beta_{\eta_2}\coloneqq\mathsf{Sam}(r)$
	
	\State  $\boldsymbol{\mathrm{u}}\coloneqq\mathsf{Compress}_{q}(\buone,d_{u})$\label{line:u}
	
	\State  $v\coloneqq\mathsf{Compress}_{q}(\boldsymbol{\mathrm{t}}^{T}\boldsymbol{\mathrm{r}}+e_2+\left \lfloor {q}/{2}\right\rfloor \cdot m,d_{v})$\label{line:v}
	
	\State \Return $c\coloneqq(\boldsymbol{\mathrm{u}},v)$

\end{algorithmic}
\end{algorithm}

\vspace{-5mm}

\begin{algorithm}[H]
\caption{${\mathsf{Kyber.CPA.Dec}}\ensuremath{(sk=\bs,c=(\bu,v))}$}
\label{alg:kyber_dec}
\begin{algorithmic}[1]

    \State
    $\bu\coloneqq\mathsf{Decompress}_{q}(\bu,d_{u})$

    \State
    $v\coloneqq\mathsf{Decompress}_{q}(v,d_{v})$

    \State \Return $\mathsf{Compress}_{q}(v-\bs^{T}\bu,1)$

\end{algorithmic}
\end{algorithm}

\vspace{-5mm}

\begin{table}[ht]
\caption{Parameters and Variants of Kyber in \cite{NISTpqcfinal2024}}
\label{Kyber_Par}\centering
\vspace{-3mm}
\begin{tabular}{|c|c|c|c|c|c|c|c|c|c|}
\hline
& $k$ & $q$ & $\eta_{1}$ & $\eta_{2}$ & $d_{u}$ & $d_{v}$ & DFR & CER\\ \hline
KYBER512 &  $2$ & $3329$ & $3$ & $2$ & $10$ & $4$ & $2^{-139}$ & $24$\\ \hline
KYBER768 & $3$ & $3329$ & $2$ & $2$ & $10$ & $4$ & $2^{-164}$ &$34$\\ \hline
KYBER1024 & $4$ & $3329$ & $2$ & $2$ & $11$ & $5$ & $2^{-174}$ &$49$ \\ \hline
\end{tabular}
\vspace{-2mm}
\end{table}

\subsection{Decryption noise, toroidal distance, and Minal codes}

The Kyber decryption process can be expressed as \cite{Kyber2021}
\begin{equation}
\setlength{\abovedisplayskip}{3pt}
\setlength{\belowdisplayskip}{3pt}
y = v - \mathbf{s}^T \mathbf{u}
= \left \lfloor {q}/{2} \right\rfloor \cdot m + n_e ,
\label{decoding_mode}
\end{equation}
where \( n_e \) denotes the decryption decoding noise. This noise term can be
expanded as
\begin{align}
n_e&= \mathbf{e}^T \mathbf{r} + e_2 + c_v
- \mathbf{s}^T \big( \mathbf{e}_1 + \mathbf{c}_u \big),
\label{Ne}
\end{align}
where \( (c_v, \mathbf{c}_u) \) denote the quantization noises introduced by the
quantization operator \( \mathsf{Compress}_{q} \) in
Algorithm~\ref{alg:kyber_enc}.

It is convenient to represent the ring elements \( m \) and \( n_e \) as column vectors of integer coefficients, i.e.,
\begin{equation}
m = [m_0, m_1, \ldots, m_{n-1}]^{T}, \; n_e = [n_0, n_1, \ldots, n_{n-1}]^{T}. \label{m_ne_vec}
\end{equation}
The entries \( n_i \) are generally dependent, as they arise from polynomial products over the ring \( R_q \) in~\eqref{Ne}.

\begin{prop}[polynomial splitting \cite{Minal_Code_Kyber2024}] \label{prop:1}
Let $\ell$, $\nu$, and $n$ be positive integers with $n$ and $\ell$ powers of $2$, $1 < \ell < n$, and $\nu = n/\ell$.  
Let $a,b \in R_q = \mathbb{Z}_q[x]/(x^n + 1)$, with coefficients sampled from distributions $\mathcal{D}_1$ and $\mathcal{D}_2$, where $\mathcal{D}_2$ is symmetric.  
Let $c = ab\in R_q$. Then, for all $0 \le i < \nu$,  
\[
\Pr\!\big( c[i, i+\nu, i+2\nu, \ldots, i+n-\nu] \big)
=
\Pr\!\big( c[0, \nu, 2\nu, \ldots, n-\nu] \big).
\]
\end{prop}

Proposition \ref{prop:1} allows us to separate the joint probability distribution of
$\ell$ coefficients from the $n$ coefficients in ${n}_e$:
\begin{equation*}
\Pr (\Delta {n}_{e}) = \Pr\!\big( [n_0, n_{\nu}, n_{2\nu}, \ldots, n_{n-\nu}] \big), 
\end{equation*}
which can be computed via the following 
corollary.

\begin{corollary}[\cite{Minal_Code_Kyber2024}]\label{coro:1}
Let \( \mathcal{P}^{(\phi_a,\phi_b)}_{\mathrm{prod}} \) denote the probability
distribution of the product \( c = a b \) of two polynomials
\( a, b \in \mathbb{Z}_q[x]/(x^\ell + 1) \), whose coefficients are sampled
according to distributions \( \phi_a \) and \( \phi_b \), respectively.
Let  $\mathcal{D}_{\mathbf{c}_u}$ denote the distribution of the coefficients of $\mathbf{c}_u$ and \( \mathcal{P}_{c_v + e_2} \) be the probability
distribution of
\( c_v + e_2\).
Then, it holds
\[
\Pr (\Delta {n}_{e}) =
\left(\mathcal{P}^{(\beta_{\eta_1}, \beta_{\eta_1})}_{\mathrm{prod}} \right)^{*\frac{k n}{\ell}}
*
\left(\mathcal{P}^{(\beta_{\eta_1},\, \beta_{\eta_2} *\mathcal{D}_{\mathbf{c}_u})}_{\mathrm{prod}}\right)^{*\frac{k n}{\ell}}
*
\mathcal{P}_{c_v + e_2}^{\otimes \ell}.
\]
\end{corollary}

Corollary~\ref{coro:1} characterizes the joint distribution of \(\ell\) decryption noise coefficients. 
Consequently, the DFR of Kyber can be upper bounded via a union bound 
over the decoding error events of all \(\ell\)-dimensional message blocks. 
Without loss of generality, let $\Delta v =  \left\lceil {q}/{2} \right\rfloor \cdot [m_0, m_{\nu}, m_{2\nu}, \ldots, m_{n-\nu}] \in \mathcal{C}$, where \(\mathcal{C} \subset \mathbb{Z}_q^{\ell}\) denotes the message codebook. 

The DFR is then upper bounded as \cite{Minal_Code_Kyber2024}
\begin{equation}
\mathrm{DFR} \le
\frac{\nu}{|\mathcal{C}|}
\sum_{\Delta v \in \mathcal{C}}
\left(
\sum_{\Delta v' \neq \Delta v \in \mathcal{C}}
p_{\mathrm{error}}(\Delta v',\Delta v), \label{DFR_Bound}
\right), \; \text{where}
\end{equation}
\begin{equation*}
p_{\mathrm{error}}(\Delta v',\Delta v)
=
\Pr\!\left(
\left\langle
\Delta n_e,\, \Delta v' - \Delta v
\right\rangle
\ge
\frac{1}{2}
\mathrm{dist}_q (\Delta v', \Delta v )^{2}
\right)
\end{equation*}
The distance function \(\mathrm{dist}_q(\cdot,\cdot)\) is defined below.

\begin{definition}[$L_2$-norm toroidal distance]
Let \(\mathbf{v}_1, \mathbf{v}_2 \in \mathbb{Z}_q^\ell\). 
The \emph{$L_2$-norm toroidal distance} is defined as
\[
\mathrm{dist}_q(\mathbf{v}_1, \mathbf{v}_2) = \left\| (\mathbf{v}_1 - \mathbf{v}_2) \bmod^\pm q \right\|,
\]
where the modulo operation \(\bmod^\pm q\) is taken component-wise, such that the result lies in \([-q/2, q/2)^\ell\), and \(\|\cdot\|\) denotes the standard Euclidean norm.  
\end{definition}

In \cite{Minal_Code_Kyber2024}, the authors propose searching for a codebook \(\mathcal{C} \subset \mathbb{Z}_q^{\ell}\) that minimizes the DFR bound in \eqref{DFR_Bound}, restricting its generator matrix to be circulant to reduce complexity.

\begin{definition}[Minal code \cite{Minal_Code_Kyber2024}] \label{def:minal}
Let \(\ell \ge 2\), \(q\) be a prime, and \(\gamma \in \mathbb{Z}\) with
\(0 \le \gamma < q/2\).
The \(\ell\)-dimensional \emph{Minal code} over \(\mathbb{Z}_q\) with parameter
\(\gamma\) is the infinite lattice
\[
\mathcal{C}_\mathrm{M}(\gamma)
=
\left\{
\mathbf{G}\mathbf{m} \bmod q
:\;
\mathbf{m} \in \mathbb{Z}_2^{\ell}
\right\},
\]
where \(\mathbf{G} \in \mathbb{Z}^{\ell \times \ell}\) is the circulant matrix
generated by
\(\bigl[\lfloor q/2 \rfloor,\, \gamma,\, 0,\, \ldots,\, 0\bigr]\).
The matrix \(\mathbf{G}\) is called the generator matrix, and \(\gamma\) is the
\emph{tailoring parameter}. 

\end{definition}

Minal codes are periodic over \(\mathbb{Z}_q^{\ell}\), but unlike linear or lattice codes, they do not form additive groups and are not closed under addition, necessitating exhaustive-search decoding. Optimizing the tailoring parameter \(\gamma\) to minimize the DFR requires prior knowledge of the joint noise distribution \(\Pr(\Delta n_e)\), which is computationally expensive as it involves convolving \(O(n)\) distinct \(\ell\)-dimensional distributions. No closed-form expression for \(\gamma\) was given in \cite{Minal_Code_Kyber2024}. Consequently, the exhaustive code search and high-complexity decoding restrict practical use of Minal codes in Kyber to \(\ell \leq 4\), leaving potential gains from longer codes unexplored. Whether codes exist that outperform the Minal code remains unknown.

\section{Maximum Toroidal Distance Code}

\begin{table*}[!t]
\centering
\caption{Exact DFR and Minimum toroidal distance of different encoding scheme}
\label{Com_M_DFR}\centering
\vspace{-3mm}
\begin{threeparttable}
\begin{tabular}{|c||c|c|c|c||c|c|c|}
\hline
\multirow{2}{*}{} & \multicolumn{4}{c||}{Literature} & \multicolumn{3}{c|}{This work} \\
\cline{2-8}
& $\mathcal{C}_\mathrm{MTD}$ \cite{NISTpqcfinal2024} \cite{KyberCode}& $\mathcal{C}_\mathrm{M}(\gamma)$ \cite{Minal_Code_Kyber2024} & $\mathcal{C}_\mathrm{M}(\gamma)$ \cite{Minal_Code_Kyber2024} & $\mathcal{C}_\mathrm{MLD}$\cite{ChiangWolf1971} & $\mathcal{C}_\mathrm{MTD}$& $\mathcal{C}_\mathrm{GTD}$ & $\mathcal{C}_\mathrm{GTD}$  \\ \hline
$\ell$&  $1$ & 2  & 4 & 4 & $2$ & $4$  & $8$  \\ \hline
$d_{\min, q}$&  $0.5q$ & 0.517q & $0.547q$ &  $0.4q$ & $0.518q$ & $0.577q$  & $0.707q$  \\ \hline
DFR: KYBER1024 ($d_u=11, d_v=5, \text{CER}=49$)&  $2^{-174}$ & $2^{-185}$ &  $2^{-201}$ & $2^{-103}$ & $2^{-185}$ & $2^{-213}$  & $2^{-286}$  \\ \hline
DFR: KYBER1024 ($d_u=10, d_v=5, \text{CER}=45$)&  $2^{-143}$  & $2^{-151}$ & $2^{-165}$& $2^{-85}$ & $2^{-152}$ & $2^{-176}$  & $2^{-239}$  \\ \hline
\end{tabular}%
\vspace{-3mm}
\end{threeparttable}
\end{table*}

We observe from \eqref{DFR_Bound} that the DFR decreases as the minimum
toroidal distance among codewords increases. This motivates the
construction of \emph{maximum $L_2$-norm toroidal distance} (MTD) codes in the
\(\ell\)-dimensional discrete torus \(\mathbb{Z}_q^{\ell}\).

\begin{definition}[MTD code]
Let \(q \ge 2\) and \(\ell \ge 1\) be integers, with $\ell$ a power of $2$. Let
\(\mathcal{C} \subset \mathbb{Z}_q^{\ell}\) be a codebook with $|\mathcal{C}|=2^{\ell}$. The \emph{minimum $L_2$-norm toroidal distance} of \(\mathcal{C}\) is defined as
\begin{equation}
\setlength{\abovedisplayskip}{3pt}
\setlength{\belowdisplayskip}{3pt}
d_{\min,q}(\mathcal{C}) = \min_{\mathbf{v}_1 \neq \mathbf{v}_2 \in \mathcal{C}} \mathrm{dist}_q(\mathbf{v}_1, \mathbf{v}_2),
\label{eq:min_toroidal_distance}
\end{equation}
A code \(\mathcal{C}_\mathrm{MTD}\) is called an
MTD code if it satisfies
\begin{equation}
\setlength{\abovedisplayskip}{3pt}
\setlength{\belowdisplayskip}{3pt}
  \mathcal{C}_\mathrm{MTD} = \arg  \max_{\mathcal{C} \subset \mathbb{Z}_q^{\ell}, |\mathcal{C}|=2^{\ell}} d_{\min,q}(\mathcal{C}).
\end{equation}
\end{definition}

\begin{example}
For $\ell = 1$, the MTD code is $\mathcal{C}_{\mathrm{MTD}} = \{0, \lfloor q/2 \rfloor\}$, as used in the original Kyber Encryption Algorithm \ref{alg:kyber_enc}.
\end{example}

The advantages of MTD codes for lattice-based encryption are twofold. 
First, they can be constructed without explicitly computing the 
DFR. Second, their construction is not restricted to circulant structures, 
enabling improved decoding performance compared to the Minal codes. Since the MTD codes are not known for any $(q,\ell)$, we introduce the following notion.

\begin{definition}[Good Toroidal Distance Code (GTD)]
A code $\mathcal{C} \subseteq \mathbb{Z}_q^\ell$ with $|\mathcal{C}| = 2^{\ell}$ is called a
\emph{good toroidal distance code} (GTD), denoted by $\mathcal{C}_{\mathrm{GTD}}$,
if it attains the largest known
$d_{\min,q}(\mathcal{C})$ among all codes with the same parameters $(q,\ell)$.
\end{definition}

We refer to MTD and GTD codes as a family of \emph{$L_2$-norm toroidal distance codes}. In the remainder of this section, we construct such codes for $\ell=2,4,8$.

\subsection{MTD code for $\ell=2$}
We first propose a variant of the Minal code that maximizes its minimum toroidal distance for any $\ell$, and then show that this variant is essentially an MTD code when $\ell=2$.

\begin{lemma}\label{lem: opt_minal}
The minimum toroidal distance of $\mathcal{C}_\mathrm{M}(\gamma)$ is
\begin{equation*}
d_{\min,q}(\mathcal{C}_\mathrm{M}(\gamma)) = \min \Big\{ \sqrt{\lfloor q/2 \rfloor^2 + \gamma^2}, \sqrt{\ell}\,(\lfloor q/2 \rfloor - \gamma) \Big\}.
\label{M_dmin}
\end{equation*}
\end{lemma}

\begin{IEEEproof}
Let $m, m' \in \mathbb{Z}_2^{\ell}$ with $m \neq m'$, and define
$m_{\Delta} = m - m' \in \{-1,0,1\}^{\ell} \setminus \{0\}$. For simplicity, denote
\[
\mathrm{dist}_q(\mathbf{G} m, \mathbf{G} m') = \|\mathbf{G} m_{\Delta}\|_{\mathrm{tor}}.
\]

\noindent\emph{Step 1: Single-bit vectors.}  
If $m_{\Delta}$ has Hamming weight $w = 1$, then $\mathbf{G} m_{\Delta}$ contains
exactly one $\pm \lfloor q/2 \rfloor$ and one $\pm \gamma$. Thus,
\[
\|\mathbf{G} m_{\Delta}\|_{\mathrm{tor}}
= \sqrt{\lfloor q/2 \rfloor^2 + \gamma^2}.
\]

\noindent\emph{Step 2: Full-weight vectors.} If $m_{\Delta}$ has Hamming weight $w=\ell$, then
$(\mathbf{G} m_{\Delta})_i = s_i \lfloor q/2 \rfloor + s_{i-1}\gamma$ for
$s_i \in \{\pm1\}$. The minimal toroidal distance occurs when the signs of $m_\Delta$ alternate as $[1,-1,1,-1,\dots]$, producing coordinate magnitudes $\lfloor q/2 \rfloor - \gamma$. Hence, for all full-weight vectors, it holds
\[
\|\mathbf{G} m_\Delta\|_{\mathrm{tor}} \geq \sqrt{\ell}\,(\lfloor q/2 \rfloor - \gamma).
\]

\noindent\emph{Step 3: Intermediate-weight vectors.}  
For $2 \le w \le \ell-1$, the circulant structure of $\mathbf{G}$ guarantees that
$\mathbf{G}m_\Delta$ has at least two nonzero coordinates whose toroidal
magnitudes are $\lfloor q/2 \rfloor$ and $\gamma$, respectively. Hence,
\[
\|\mathbf{G} m_\Delta\|_{\mathrm{tor}}
\ge \sqrt{\lfloor q/2 \rfloor^2 + \gamma^2}.
\]

\noindent\emph{Step 4:}  
Comparing all cases establishes the claim.
\end{IEEEproof}

\begin{prop}[Optimal $\gamma$] \label{prop: opt_minal}
For $\gamma \in [0,q/2)$, $d_{\min,q}(\mathcal{C}_\mathrm{M}(\gamma))$ is maximized at the unique interior point
\begin{equation}
\gamma^\star 
= \frac{\lfloor q/2 \rfloor (\ell - \sqrt{2\ell-1})}{\ell-1}.
\label{gamma_opt_updated}
\end{equation}
The corresponding maximum minimum toroidal distance is
\begin{equation*}
d_{\min,q}(\mathcal{C}_\mathrm{M}(\gamma^\star)) = \sqrt{\lfloor q/2 \rfloor^2 + (\gamma^\star)^2} 
= \sqrt{\ell}\,(\lfloor q/2 \rfloor - \gamma^\star).
\end{equation*}
\end{prop}

\begin{IEEEproof}
$d_{\min,q}(\mathcal{C}_\mathrm{M}(\gamma))$ is the minimum of two terms:
\[
f_1(\gamma) = \sqrt{\lfloor q/2 \rfloor^2 + \gamma^2}, \qquad 
f_2(\gamma) = \sqrt{\ell}\,(\lfloor q/2 \rfloor - \gamma).
\]

To maximize the minimum, we equate the two terms:
\[
\sqrt{\lfloor q/2 \rfloor^2 + \gamma^2} = \sqrt{\ell}\,(\lfloor q/2 \rfloor - \gamma).
\]
Squaring and rearranging gives a quadratic in $\gamma$:
\[
(\ell-1) \gamma^2 - 2 \ell \lfloor q/2 \rfloor \, \gamma + \lfloor q/2 \rfloor^2 (\ell - 1) = 0.
\]

Solving yields a unique root in $(0,q/2)$, given by \eqref{gamma_opt_updated}.  
Since $f_1(\gamma)$ is strictly increasing and $f_2(\gamma)$ is strictly decreasing on $[0,q/2)$, this intersection point uniquely maximizes $\min\{f_1,f_2\}$.  
At $\gamma = \gamma^\star$, the two terms are equal, giving the maximum minimum toroidal distance.
\end{IEEEproof}

The following theorem shows that the optimized Minal code $\mathcal{C}_M(\gamma^\star)$ in Proposition \ref{prop: opt_minal} is, in fact, an MTD code for $\ell=2$.

\begin{theo}[MTD code for $\ell=2$]
\label{thm:MTD_l2}
Let $a = \lfloor q/2 \rfloor$. Any 4-point code $\mathcal{C} \subset \mathbb{Z}_q^2$ admits an MTD-optimal code $\mathcal{C}_{\mathsf{MTD}}$ that, up to torus isometry and translation, can be written as
\[
\mathcal{C}_{\mathsf{MTD}} = \{(0,0), (a,\gamma), (\gamma,a), (a+\gamma,a+\gamma)\}, \quad \gamma \in [0,a),
\]
and the optimal code coincides with the 2-D Minal code $\mathcal{C}_\mathrm{M}(\gamma^\star)$, where $\gamma^\star$ is given by Proposition~\ref{prop: opt_minal}.
\end{theo}

\begin{IEEEproof}[Proof sketch]
The proof proceeds in four steps:

\noindent \emph{Step 1: Reduction to a parallelogram.}  
By translation invariance of the toroidal distance, we may assume
$0 \in \mathcal{C}$.
Let $\mathbf{u},\mathbf{v}$ be two codeword differences attaining the minimum
toroidal distance.
If the diagonal $\mathbf{u}+\mathbf{v}$ were closer to the origin than
$\mathbf{u}$ or $\mathbf{v}$, one could shorten $\mathbf{u}$ or $\mathbf{v}$
and rebalance the configuration to strictly increase the minimum distance,
contradicting optimality.
Hence, it suffices to consider codes of the parallelogram form
$\{0,\mathbf{u},\mathbf{v},\mathbf{u}+\mathbf{v}\bmod q\}$.

\noindent \emph{Step 2: Parameterization.}  
Inside the fundamental cube $[0,a]^2$, write
$\mathbf{u} = (u_1,u_2), \quad \mathbf{v} = (v_1,v_2), \quad 0 \le u_i, v_i \le a$, where $a = \lfloor q/2 \rfloor$.  
This is always possible by choosing canonical representatives modulo $q$, which preserves all pairwise toroidal distances.  
The six pairwise distances among the four points are exactly Euclidean in the cube, so the optimization reduces to maximizing the minimum of these six distances over $(u_1,u_2,v_1,v_2)\in[0,a]^4$.

\noindent \emph{Step 3: Symmetrization.}  
Saturating the largest coordinates ($u_1 \allowbreak = v_2 \allowbreak = a$) and setting the remaining free coordinates equal ($u_2 \allowbreak = v_1 \allowbreak = \gamma$) preserves or improves the minimum distance due to symmetry.  
The parallelogram then depends on a single parameter $\gamma \in [0,a)$:
\[
\mathbf{u}=(a,\gamma), \mathbf{v}=(\gamma,a),  
\mathcal{C} = \{(0,0),(a,\gamma),(\gamma,a),(a+\gamma,a+\gamma)\}.
\]

\noindent \emph{Step 4: Identification as Minal code.}  
This one-parameter family coincides with the 2-dimensional Minal code. Maximizing the minimum toroidal distance over $\gamma$ yields $\gamma^\star$ from Proposition~\ref{prop: opt_minal}.  
Hence, the optimal MTD code is 
\begin{equation*}
\mathcal{C}_\mathrm{M}(\gamma^\star) = \{(0,0),(a,\gamma^\star),(\gamma^\star,a),(a+\gamma^\star,a+\gamma^\star)\}. \qquad \qquad \IEEEQEDhere
\end{equation*}
\end{IEEEproof}

\begin{example}\label{Ex_M}
For $q=3329$, $\mathcal{C}_M(\gamma^\star)$ yields
\[
\begin{array}{c|c|c}
\ell & \gamma^\star & d_{\min,q}(\mathcal{C}_M(\gamma^\star))/q \\ \hline
2  & \approx 446 & \approx 0.518 \\
4  & \approx 751 & \approx 0.548 \\
8  & \approx 981 & \approx 0.580 \\
16 & \approx 1157 & \approx 0.609
\end{array}
\]
\end{example}

We observe that, for optimized Minal codes, the minimum toroidal distance grows only slowly with the dimension~$\ell$. This limitation is inherent to the circulant structure of the generator matrix, which constrains geometric diversity. Consequently, optimized Minal codes are generally not MTD codes for $\ell>2$.

\subsection{GTD code construction via lattices for any $\ell$}\label{subsec: GTD}
This subsection presents a generic GTD code construction, which is subsequently specialized to \(\ell = 2\) and \(\ell = 8\), yielding a larger minimum $L_2$-norm toroidal distance than the optimized Minal codes in Example~\ref{Ex_M}.

To simplify the construction, we first build a code $\mathcal{C}'$ over a smaller space $\mathbb{Z}_p^\ell$ with $p<q$, and then scale it to $\mathbb{Z}_q^\ell$:
\begin{equation}
    \mathcal{C}_\mathrm{GTD} = \lfloor {q}/{p} \rfloor \, \mathcal{C}' \subset \mathbb{Z}_q^\ell.
\end{equation}
The lemma below enables GTD code construction from lattice.

\begin{lemma}\label{lem:d_min}
Let $\Lambda \subset \mathbb{R}^\ell$ be a lattice containing the sublattice $p\,\mathbb{Z}^\ell$.  
Its minimum Euclidean distance is
\begin{equation}
d_{\min}(\Lambda) = \min_{\mathbf{v}_1 \neq \mathbf{v}_2 \in \Lambda} \|\mathbf{v}_1 - \mathbf{v}_2\| = d_{\min,p}(\Lambda \cap \mathbb{Z}_p^\ell). \label{ED_TD}    
\end{equation}
\end{lemma}

\vspace*{1pt}
\begin{IEEEproof}
For $\mathbf{v}_1, \mathbf{v}_2 \in \Lambda \cap \mathbb{Z}_p^\ell$, we have
$\mathrm{dist}_p(\mathbf{v}_1, \mathbf{v}_2) = \min_{\mathbf{k} \in p\mathbb{Z}^\ell} \|\mathbf{v}_1 - \mathbf{v}_2 + \mathbf{k}\|$. 
Since $p\mathbb{Z}^\ell \subset \Lambda$, all candidates lie in $\Lambda$, implying
$d_{\min,p}(\Lambda \cap \mathbb{Z}_p^\ell) \ge d_{\min}(\Lambda)$.

Conversely, let $\mathbf{w}\in\Lambda$ attain $d_{\min}(\Lambda)$.
Reducing $\mathbf{w}$ modulo $p\mathbb{Z}^\ell$ yields $\mathbf{v}\in\Lambda\cap\mathbb{Z}_p^\ell$
with $\|\mathbf{v}\|=\|\mathbf{w}\|$.
Thus $\mathrm{dist}_p(\mathbf{0},\mathbf{v})=d_{\min}(\Lambda)$, and
$d_{\min,p}(\Lambda\cap\mathbb{Z}_p^\ell)\le d_{\min}(\Lambda)$.
\end{IEEEproof}

Lemma~\ref{lem:d_min} reduces code construction over $\mathbb{Z}_p^\ell$ with large toroidal distance to finding a lattice $\Lambda$ containing the sublattice $p\mathbb{Z}^\ell$ with a large minimum Euclidean distance. Thus, the codebook $\mathcal{C}'$ corresponds to the lattice points in $\Lambda \cap \mathbb{Z}_p^\ell$. Recall that both MTD and GTD codes have size $2^\ell$. The following lemma provides a simple method to count the points in $\Lambda \cap \mathbb{Z}_p^\ell$.

\begin{lemma}[\cite{BK:Conway93}]\label{lem:size_of_C}
Let $\Lambda \subset \mathbb{R}^\ell$ be a lattice containing the sublattice $p\,\mathbb{Z}^\ell$.  
Then the number of points in $\Lambda \cap \mathbb{Z}_p^\ell$ is
\[
N=|\Lambda \cap \mathbb{Z}_p^\ell| = {p^\ell}/{\det(\Lambda)},
\]
where $\det(\Lambda)$ denotes the determinant of $\Lambda$.
\end{lemma}

\emph{GTD code construction via lattices for any $\ell$:}  
Based on Lemmas~\ref{lem:d_min} and~\ref{lem:size_of_C}, we first select a lattice $\Lambda$ containing the sublattice $p\,\mathbb{Z}^\ell$ that has the largest known minimum Euclidean distance.  
From $\Lambda \cap \mathbb{Z}_p^\ell$, we select $2^\ell$ lattice points out of the total $N$ lattice points that maximize the minimum pairwise toroidal distance, forming the codebook $\mathcal{C}'$. 
This construction ensures $d_{\min,p}(\mathcal{C}') \geq d_{\min}(\Lambda)$, or equivalently,
\[
d_{\min,q}(\mathcal{C_\mathrm{GTD}}) = d_{\min,q}(\lfloor {q}/{p} \rfloor \, \mathcal{C}') \geq \lfloor {q}/{p} \rfloor d_{\min}(\Lambda).
\]
Equality holds for $N=2^{\ell}$.

\subsection{GTD code for $\ell=4$}
For $\ell=4$,  the $D_4$ lattice provides the largest known lattice coding gain, denoted as $r(\Lambda) = {d_{\min}^2(\Lambda)}/{\det(\Lambda)^{2/\ell}}$, making it the best candidate to construct $\mathcal{C}_\mathrm{GTD}$. The $D_4$ lattice has determinant $2$ and contains the sublattice $p\mathbb{Z}^4$ for any even $p>0$, setting $p=6$, we obtain the codebook below.

\begin{example}[GTD code for $\ell=4$] \label{eg_C4}
For $\ell=4$, the codebook $\mathcal{C}_{\mathrm{GTD}}$ achieves
$d_{\min,q}(\mathcal{C}_{\mathrm{GTD}}) \approx 0.577\,q$, outperforming the optimized Minal codes in Example~\ref{Ex_M}.  
The codebook is
\[
\mathcal{C}_{\mathrm{GTD}} = \left\lfloor \dfrac{q}{6} \right\rfloor \cdot
\left\{
\begin{aligned}
&(0,0,0,0),\;(4,2,4,0),\;(3,3,3,3),\;(2,0,4,2),\\
&(2,4,2,0),\;(3,1,1,1),\;(1,1,3,5),\;(1,5,1,3),\\
&(0,2,2,2),\;(3,5,5,5),\;(1,3,5,1),\;(0,4,4,4),\\
&(4,0,2,4),\;(4,4,0,2),\;(5,1,5,3),\;(5,5,3,1)
\end{aligned}
\right\}
\]
\end{example}

\subsection{MTD and GTD codes for $\ell=8$}
We first construct an MTD code over $\mathbb{Z}_4^8$, then scale it to obtain a GTD code over $\mathbb{Z}_q^8$.
\begin{theo}[MTD code in $\mathbb{Z}_4^8$]\label{theo:MTD_z_4}
For any code $\mathcal{C}' \subset \mathbb{Z}_4^8$ with $|\mathcal{C}'| = 2^8$, the minimum $L_2$-norm toroidal distance satisfies
\[
d_{\min,4}(\mathcal{C}') \le d_{\min}(2E_8) = 2 \sqrt{2}.
\]
Equality holds for $\mathcal{C}' = 2E_8 \cap \mathbb{Z}_4^8$, the MTD code in $\mathbb{Z}_4^8$.
\end{theo}

\begin{IEEEproof}
Let $\mathcal{P}(\mathcal{C}') = \mathcal{C}' + 4\mathbb{Z}^8 \subset \mathbb{R}^8$ denote the periodic extension of $\mathcal{C}'$.  
Each fundamental cell of $4\mathbb{Z}^8$ contains exactly $2^8$ points, so the \emph{per-point volume} is $4^8 / 2^8 = 2^8$. By definition,
\[
d_{\min}(\mathcal{P}(\mathcal{C}')) = \min_{v_1 \neq v_2 \in \mathcal{C}',\, k \in 4\mathbb{Z}^8} \|v_1 - v_2 + k\| = d_{\min,4}(\mathcal{C}').
\]

\vspace*{1pt}
The scaled lattice $\Lambda = 2E_8$ has $\det(\Lambda) = 2^8$ \cite{BK:Conway93}, matching the per-point volume of $\mathcal{P}(\mathcal{C}')$. Viazovska~\cite{Viazovska2017} shows that among all periodic point sets in $\mathbb{R}^8$ with volume $2^8$ per point, $\Lambda$ achieves the largest minimum Euclidean distance. Hence
\[
d_{\min,4}(\mathcal{C}') = d_{\min}(\mathcal{P}(\mathcal{C}')) \le d_{\min}(\Lambda) = 2\sqrt{2}.
\]

Consider $\mathcal{C}' = \Lambda \cap \mathbb{Z}_4^8$. By Lemma~\ref{lem:size_of_C}, $\Lambda \supset 4\mathbb{Z}^8$~\cite{shuiyin2024}, and $\det(\Lambda)=2^8$, giving $|\mathcal{C}'| = |\Lambda \cap \mathbb{Z}_4^8| = 4^8/2^8 = 2^8$.  Lemma~\ref{lem:d_min} yields $d_{\min,4}(\mathcal{C}') = d_{\min}(\Lambda)$, so equality is attained.
\end{IEEEproof}

\begin{corollary}[GTD Code in $\mathbb{Z}_q^8$]
The scaled codebook
\[
\mathcal{C}_{\mathsf{GTD}} = \left\lfloor {q}/{4} \right\rfloor \cdot (2E_8 \cap \mathbb{Z}_4^8) \subset \mathbb{Z}_q^8
\]
forms a GTD code in $\mathbb{Z}_q^8$ with 
$d_{\min,q}(\mathcal{C}_{\mathsf{GTD}})  = 2\sqrt{2}\left\lfloor {q}/{4} \right\rfloor$.
\end{corollary}

\subsection{Comparison and decoding complexity}
In Table~\ref{Com_M_DFR}, we compare the DFRs of different encoding schemes under the Kyber-1024 parameter set, where the DFRs are numerically evaluated using~\eqref{DFR_Bound}. As the code dimension $\ell$ increases, the proposed GTD codes achieve substantially lower DFRs than existing schemes, owing to their larger $d_{\min,q}$. The values of $d_{\min,q}$ for Minal codes are computed from the corresponding $\gamma$ values reported in \cite[Table~5]{Minal_Code_Kyber2024} using Lemma~\ref{lem: opt_minal}. The reduced DFR enables tighter compression parameters $(d_u, d_v)$, resulting in a lower CER. 

In the literature, the code most closely related to MTD codes is the \emph{maximum Lee distance (MLD) code}. 
The MLD code aims at maximizing the $L_1$-norm toroidal distance, 
whereas the proposed MTD codes target maximizing the $L_2$-norm toroidal distance. 
For comparison, we consider the $(4,2,4)_5$ MLD code \cite{ChiangWolf1971} with $p=5$, $
\ell=4$, and generator matrix
\[
\mathbf{G}_\mathrm{MLD} =
\begin{pmatrix}
3 & 4 & 1 & 0 \\
0 & 3 & 4 & 1
\end{pmatrix} \in \mathbb{Z}_5^{2 \times 4}.
\]
This code contains $25$ codewords and achieves minimum Lee distance $4$. 
From these, we select $16$ codewords with the largest minimum Lee distance, scale them by $\left\lfloor {q}/{5} \right\rfloor$, and report their DFR in Table~\ref{Com_M_DFR}. We observe that the obtained MLD code $\mathcal{C}_\mathrm{MLD}$ exhibits poor DFR performance compared with other codes due to its smaller minimum $L_2$-norm toroidal distance, confirming that the $L_2$-norm toroidal distance provides a more accurate measure of reliability in lattice-based encryption.

The decoding complexity of the proposed codes is negligible for $\ell \le 4$, even under exhaustive search, compared with decryption. For $\ell=8$, a fast constant-time closest-vector decoder for the $E_8$ lattice is available \cite{SloaneDn1982}.

\section{Conclusions}

We proposed a family of codes for lattice-based encryption schemes that maximize the minimum pairwise $L_2$-norm toroidal distance among $2^{\ell}$ codewords over the $\ell$-dimensional torus $\mathbb{Z}_q^\ell$. These codes are closely related to lattices containing the sublattice $q\mathbb{Z}^\ell$. We illustrated constructions via the Minal code ($\ell=2$), the $D_4$ lattice ($\ell=4$), and the $E_8$ lattice ($\ell=8$). Compared with existing schemes, the proposed codes achieve substantially lower DFRs. An open problem remains the construction of fast-decodable MTD codes for $\ell>8$.

\bibliographystyle{IEEEtran}
\bibliography{IEEEabrv,LIUBIB}

\end{document}